\documentclass[12pt,letterpaper]{article}%
\usepackage{amsmath,amssymb,amsfonts}
\usepackage{caption2}
\usepackage[]{hyperref}
\usepackage[pdftex]{color,graphicx}

\newcommand{\hhref}[1]{\href{http://arxiv.org/abs/#1}{arXiv:#1}}
\numberwithin{equation}{section} 
\setcaptionmargin{1cm}

\begin{document}

\begin{titlepage}
\begin{flushright}
\end{flushright}
\vskip 1.0cm
\begin{center}
{\Large \bf Naturalness bounds in extensions of the MSSM \\
 without a light Higgs boson} \vskip
1.0cm {\large Paolo Lodone} \\[0.7cm]
{\it Scuola Normale Superiore and INFN, Piazza dei Cavalieri 7, 56126 Pisa, Italy} \vskip 2.0cm
\end{center}
\begin{abstract}

Adopting a bottom-up point of view, we make a comparative study of the simplest
extensions of the MSSM with extra tree level contributions to the lightest Higgs boson mass.
We show to what extent a relatively heavy Higgs boson, up to 200-350 GeV, can be compatible with data and naturalness. The price to pay is that the theory undergoes some change of regime at a relatively low scale.
Bounds on these models come from electroweak precision tests and naturalness, which often requires the scale at which the soft terms are generated to be relatively low.

\vskip 1.5cm April 2010
\end{abstract}

\end{titlepage}

\section{Introduction and motivations}

The unexplained large difference between the Fermi scale and the Planck scale is the main reason why the Standard Model Higgs sector is widely held to be incomplete. Low energy Supersymmetry provides one of the most attractive solutions to this hierarchy.
Its main virtues of are, virtually: i) naturalness, ii) compatibility with Electroweak Precision Tests (EWPT),
iii) perturbativity, and iv) manifest unification.
However, after the LEP2 bound $m_h > 114.4$ GeV \cite{Barate:2003sz} on the lightest Higgs boson mass, the MSSM has a serious problem in dealing with (i). The reason is that $m_h$ cannot exceed $m_Z$ at tree level, and increasing it through large radiative corrections goes precisely in the direction of unnaturalness. Even with the addition of extra matter in the loops \cite{Martin:2009bg} it is difficoult to go much beyond $115$ GeV without a large amount of finetuning.
This motivates the study of models with extra tree level contributions to the Higgs quartic coupling, and thus to the masses of the Higgs sector. 

There can be extra $F$ terms, like in the Next to Minimal Supersymmetric Standard Model (NMSSM) \cite{Fayet:1974pd}\cite{Ellis:1988er}\cite{Drees:1988fc}, or extra $D$ terms if the Higgs shares new gauge interactions \cite{Haber:1986gz}\cite{Espinosa:1991gr}\cite{Espinosa:1998re}\cite{Batra:2003nj}\cite{Delgado:2004pr}\cite{Maloney:2004rc}, or both ingredients \cite{Batra:2004vc}\cite{Babu:2004xg}.
The usual/earlier approach is focussing on unification and requiring that it be not disturbed by the extra matter and interactions: at least the new couplings must not become strong before $M_{GUT}$. For this reason it is typically difficoult to go beyond $m_h =150$ GeV.
The issue is particularly relevant in view of the LHC: should we throw away low energy Supersymmetry if the lightest Higgs boson is not found below 150 GeV?
As originally suggested by \cite{Harnik:2003rs}, the request of {\itshape manifest} unification could be highly too restrictive. In fact there can be anything between the Fermi scale and the unification scale, and we cannot conclude that unification is spoiled just because some couplings become strong at an intermediate scale.
Moreover there are explicit examples \cite{Harnik:2003rs}\cite{Chang:2004db}\cite{Birkedal:2004zx}\cite{Delgado:2005fq} in which such a change of regime indeed takes place and is consistent with unification!

Given our ignorance of the high energy behaviour of the theory and the lack of conclusive hints, we stick to a bottom-up point of view, as in \cite{Barbieri:2006bg}.
In a minimalistic approach, we focus on the simplest possible extensions of the MSSM which meet the goal: adding a new $U(1)$ or $SU(2)$ gauge interaction \cite{Batra:2003nj}, or adding a gauge singlet with large coupling to the Higgses \cite{Barbieri:2006bg}.
The only constraints come from naturalness and EWPT. In other words, we prefer to retain the virtues (i), (ii), and (iii) at low energies at the price of (iv), instead of insisting on (iv) paying the price of (i).
An alternative approach could be to insist only on (i) and (ii), giving up both (iii) and (iv) ie turning to the possibility of strongly coupled theories. In this respect, one could say that the true virtue of low energy Supersymmetry is to address (i) and (ii) while retaining (iii).

This work is organized as follows: in Section \ref{sect:U1} we consider adding a new $U(1)$ gauge group to the MSSM, in Section \ref{sect:SU2} a new $SU(2)$, and in Section \ref{sect:lambdasusy} a gauge singlet.
We then conclude in Section \ref{sect:conclusion}.
The main purpose is to give a comparative study of the simplest extensions of the MSSM proposed in the literature to accommodate for a lightest Higgs boson significantly heavier than usual, in the 200-300 GeV range of masses.
We tolerate a finetuning of 10 \%, or $\Delta=10$ according to the usual criterion \cite{Barbieri:1987fn}.
We call $\Lambda$ the scale of semiperturbativity, at which some expansion parameter becomes equal to 1, and $M$ the scale at which the soft breaking terms are generated. We will see that they are often required to be both relatively low.

A unified viewpoint on the Higgs mass and the flavor problems for this kind of models will be presented in a separate work \cite{Barbieri:2010pd}.

\section{Gauge extension $U(1)$} \label{sect:U1}

This model is proposed in \cite{Batra:2003nj} just as a warm up for the nonabelian case, and then quickly discarded.
Adopting the point of view outlined in the Introduction, we take it seriously as a simple and effective possibility.

\subsection{Description of the model}

Starting from the MSSM with right handed neutrinos, the extra ingredients are a new gauge group $U(1)_x$ associated with $T_3^R = Y + \frac{L-B}{2}$, two scalars $\phi$ and $\phi^c$ with opposite charges $\pm q$, and a singlet $s$. The charged fields are shown in Table \ref{cariche}.

\begin{table}[thb]
\begin{center}
\begin{tabular}{c|c|c|c|c|c|c|c|c|c|c}
 & $\phi$ & $\phi^c$ & $H_u$ & $H_d$ & $d$ & $u$ & $Q$ & $e$ & $n$ & $L$  \\ \hline
$Y$ & 0 & 0 & $\frac{1}{2}$ & $-\frac{1}{2}$ & $\frac{1}{3}$ & $-\frac{2}{3}$ & $\frac{1}{6}$ & 1 & 0 & $-\frac{1}{2}$ \\ \hline
$X = \frac{L-B}{2} + X_\phi$ & $q$ & $-q$ & 0 & 0 & $\frac{1}{6}$ & $\frac{1}{6}$ & $-\frac{1}{6}$ & $-\frac{1}{2}$ & $-\frac{1}{2}$ & $\frac{1}{2}$ \\ \hline 
$ Y + X $ & $q$ & $-q$ & $\frac{1}{2}$ & $-\frac{1}{2}$ & $\frac{1}{2}$ & $-\frac{1}{2}$ & 0 & $\frac{1}{2}$ & $-\frac{1}{2}$ & 0 \\ \hline
\end{tabular}
\end{center}
\caption{Charge of the various fields under $U(1)_x$.}
\label{cariche}
\end{table}

The new superpotential term:
$$
W = \lambda \, s \, (\phi \phi^c - w^2)
$$
together with the soft breaking terms:
\begin{equation} \label{softterms}
\mathcal{L}_{soft} = -M_s^2 |s|^2 - M_{(\phi)}^2 |\phi|^2 - M_{(\phi^c)}^2 |\phi^c|^2 - M_\chi  \tilde{\chi} \tilde{\chi} + B_s (\phi \phi^c + h.c.) \, ,
\end{equation}
where $\tilde{\chi}$ is the new gaugino, produce the scalar potential:
$$
V = V_{MSSM} + V_{H\phi} + V_\phi
$$
with:
\begin{eqnarray*}
V_{MSSM} &=& \mu_u^2 |H_u|^2 + \mu_d^2 |H_d|^2 + \mu_3^2 (H_u H_d + h.c.) \\
&& +\frac{1}{2} g^2 \sum_a \left( \sum_i H_i^* T^a H_i + ..\right)^2 + \frac{1}{2} g^{\prime 2}\left(  \frac{1}{2} |H_u|^2 - \frac{1}{2} |H_d|^2 + ..\right)^2 \\
V_{H\phi} &=& \frac{1}{2} g_x^2 \left( \frac{1}{2} |H_u|^2 - \frac{1}{2} |H_d|^2 + q |\phi|^2 - q |\phi^c|^2 + ..\right)^2  \\
V_\phi &=& \lambda^2 |\phi|^2 |\phi^c|^2 - B (\phi \phi^c + h.c.) + M_{(\phi)}^2 |\phi|^2 + M_{(\phi^c)}^2 |\phi^c|^2 \, .
\end{eqnarray*}
We wrote only the Higgs and $\phi$ fields in the $D$ terms.
The parameters $\mu_3^2$ and $B$ have been made real and positive through field phase redefinition.
The full interaction Lagrangian of the new sector, apart from the $D$ terms, is:
\begin{eqnarray}
\mathcal{L}_{int} &=& - \frac{1}{2} \left[ \lambda s \, \tilde{\phi} \tilde{\phi^c} + \lambda \phi \, \tilde{s} \tilde{\phi^c} + \lambda \phi^c \, \tilde{s} \tilde{\phi} + h.c. \right] - \lambda^2 |\phi \, \phi^c|^2 - \lambda^2 |s \, \phi|^2 \, \label{interactionlagrangian}  \\
&&    - \lambda^2 |s\, \phi^c|^2 + \lambda w^2 (\phi \phi^c + h.c.) -\sqrt{2} g_x q \left [\phi^{*} \, \tilde{\phi} \tilde{\chi} - \phi^{c *} \, \tilde{\phi^c} \tilde{\chi} + h.c.  \right] \, . \nonumber
\end{eqnarray}
The $B$ term in the potential $V_\phi$ has in general a soft component $B = \lambda w^2 + B_s$.
The field $s$ will be generically assumed to be heavy for our considerations.

It is easy to see that the conditions for stability and unbroken Electromagnetism and CP at tree level in the Higgs sector are basically the same as in the MSSM:
$$
\left\{
\begin{array}{l}
2 \mu_3^2 < \mu_u^2 + \mu_d^2 \\
\mu_3^4 > \mu_u^2 \mu_d^2 \,\\
\lambda^2 >0 .
\end{array}
\right.
$$
We can then write the configuration of the fields at the minimum as:
$$
H_u = \left( \begin{array}{c} 0 \\ v_u \end{array}\right) \quad , \quad H_d = \left( \begin{array}{c} v_d \\ 0 \end{array}\right) \quad , \quad \phi = u_1 \quad , \quad \phi^c = u_2 \quad
$$
with $v_i, u_i \geq 0 $, and the scalar potential reduces to:
\begin{eqnarray}
V &=& \mu_u^2 v_u^2 + \mu_d^2 v_d^2 - 2 \mu_3^2 v_u v_d  + \frac{1}{8}(g^2 + g^{\prime 2}) [v_u^2 - v_d^2]^2  \label{potentialvev} \\
&& + \frac{1}{8} g_x^2 [ v_u^2 - v_d^2 - 2q u_2^2 + 2q u_1^2 ]^2  \nonumber \\
&& + \lambda^2 u_1^2 u_2^2 - 2 B u_1 u_2  + M_{(\phi)}^2 u_1^2 + M_{(\phi^c)}^2 u_2^2 \, . \nonumber
\end{eqnarray}
Notice that the mass of the new gauge boson $Z'$ of $U(1)_x$ is:
$$
M_{Z'}^2 = 2 g_x^2 \left( q^2 (u_1^2 + u_2^2) + \frac{v_u^2 + v_d^2}{4} \right)
$$
which has to be significantly heavier than that of the weak gauge bosons. Thus we assume:
$$
u_1^2, u_2^2 \gg v_u^2, v_d^2 \, \, .
$$ 
Furthermore let us assume that the mass splitting of $\phi,\phi_c$ is small:
$$
M_{(\phi)}^2 = M_\phi^2 + \frac{\Delta M_\phi^2}{2} \quad , \quad M_{(\phi^c)}^2 = M_\phi^2 - \frac{\Delta M_\phi^2}{2} \quad , \quad \Delta M_\phi^2 \ll M_\phi^2 \, \, .
$$
Then we can look for an approximate solution of the form:
$$
<\phi> = u_1 = u + \alpha \quad , \quad <\phi^c> = u_2 = u -\alpha \quad , \quad \alpha \ll u \, .
$$
Solving perturbatively one obtains, at lowest order:
\begin{eqnarray}
u^2 &=& \frac{B - M_\phi^2}{\lambda^2} =  \frac{B_s + \lambda w^2 - M_\phi^2}{\lambda^2} \qquad (\Rightarrow \mbox{ must be: } B > M_\phi^2)  \label{newbreakingscale}\\
\alpha &=& \frac{1}{u} \, \, \frac{-qg_x^2(v_u^2 - v_d^2) - 2 \Delta M_\phi^2}{8q^2g_x^2 + \frac{4 M\phi^2}{B-M_\phi^2}\lambda^2}  \, . \nonumber
\end{eqnarray}
Substituting in (\ref{potentialvev}) and minimizing in $v_u$ and $v_d$ one finds exactly the same equations as in the MSSM, but for the replacements:
\begin{equation} \label{sostituzioni}
m_z^2  \longrightarrow  m_z^2 +  \frac{g_x^2 v^2}{2(1+\frac{M_{Z'}^2}{2 M_\phi^2})} \, , \quad
\mu_u^2  \longrightarrow  \mu_u^2 - \frac{\frac{\Delta M_\phi^2}{2q}}{1+\frac{2 M_\phi^2}{M_{Z'}^2}} \, , \quad
\mu_d^2  \longrightarrow  \mu_d^2 + \frac{\frac{\Delta M_\phi^2}{2q}}{1+\frac{2 M_\phi^2}{M_{Z'}^2}}
\end{equation}
with $M_{Z'}^2 \approx 4 q^2 g_x^2 u^2$. Notice that the first one coincides with equations (2.4) and (2.5) of \cite{Batra:2003nj}.
Making the substitutions (\ref{sostituzioni}) in the usual results we find the same expression for $\tan \beta$, while the equation relating $m_z$ to the vevs gets modified:
\begin{eqnarray}
\tan \beta &=& \frac{1}{2\mu_3^2} \left( \mu_u^2 + \mu_d^2 - \sqrt{(\mu_u^2 + \mu_d^2)^2 -4\mu_3^4} \right) \label{tanbeta} \\
m_z^2 +  \frac{g_x^2 v^2}{2(1+\frac{M_{Z'}^2}{2 M_\phi^2})} &=& \frac{\left| \mu_d^2 - \mu_u^2 + \frac{\Delta M_\phi^2 / q}{1+ 2 M_\phi^2 / M_{Z'}^2} \right|}{\sqrt{1 - \sin^2 2\beta}} - \mu_u^2 - \mu_d^2 \, .\label{mzvevs}
\end{eqnarray}
In the limit of large $\tan \beta$, assuming as usual $ \mu_d^2 > \mu_u^2 - \frac{\Delta M_\phi^2 / q}{1+ 2 M_\phi^2 / M_{Z'}^2}$, one finds:
\begin{equation} \label{mzvevsLargeTanBeta}
-\mu_u^2 = - m_{H_u}^2 - \mu^2 = \frac{m_z^2}{2} +  \frac{g_x^2 v^2}{4(1+ M_{Z'}^2 / 2 M_\phi^2)} - \frac{\Delta M_\phi^2 / q}{2(1+ 2 M_\phi^2 / M_{Z'}^2)} \, .
\end{equation}
The usual bound on the Higgs boson mass at tree level becomes:
\begin{equation} \label{boundHiggsmass}
m_h^2 \leq \left( m_z^2 +  \frac{g_x^2 v^2}{2(1+\frac{M_{Z'}^2}{2 M_\phi^2})} \right) \cos^2 2\beta \, .
\end{equation}
We will call $m_h^{max}$ the expression in brackets, which corresponds to $m_h$ at tree level for large $\tan \beta$.
Notice that the $D$ term decouples for small $\lambda$, because this means large $M_{Z'}$.
From (\ref{boundHiggsmass}) we immediatly see why the $D$ term may not decouple. The extra contribution is small in the limit of large $Z'$ mass, which is what required by EWPT. Nevertheless it remains relevant if the soft mass $M_\phi$ is large too. Thus a price to pay is that we need different soft mass scales in the theory, since $M_\phi$ has to be around $10$ TeV, as we shall see.

After the symmetry breaking, in the $\phi,\phi^c$ sector we have the massless Goldstone boson plus three real scalars with masses:
$$
\sqrt{2} M_\phi \quad , \quad \sqrt{2} \sqrt{B} \quad , \quad \sqrt{2} \sqrt{B - M_\phi^2}
$$
so that there is no problem of new light particles.
The $s$ scalar keeps its soft mass $M_s$ of (\ref{softterms}).
On the other hand from (\ref{softterms}) and (\ref{interactionlagrangian}) we see that the fermion mass matrix has eigenvalues:
$$
\pm \frac{1}{\sqrt{2}} \lambda w \quad , \quad \frac{1}{2} \left( M_\chi \pm \sqrt{M_\chi^2 + 16 g_x^2 q^2 w^2} \right) \, .
$$
Thus $\lambda$ cannot be too small otherwise there are light fermions in the spectrum. The soft parameter $M_\chi$ instead can be small, since this will not correspond to light particles.

\subsection{Naturalness bounds}

Since the largest possible Higgs boson mass is realized for $M_{Z'}^2 \ll M_\phi^2$, we have to worry about finetunig at tree level in the potential $V_\phi$. Naturalness of the scale $u$ means that it must be:
$$
\Delta_u = \left| \frac{\partial \log u^2}{\partial \log M_\phi^2} \right| = \frac{M_\phi^2}{B - M_\phi^2} \leq 10
$$
so that:
\begin{equation} \label{boundOnMphi}
\frac{M_{Z'}^2}{2 M_\phi^2} = \frac{2 q^2 g_x^2}{\lambda^2} \frac{1}{\Delta_u} \geq \frac{1}{5 \lambda^2} q^2 g_x^2 \, .
\end{equation}
On the other hand from (\ref{mzvevs}) we obtain, in the limit of large $\tan \beta$:
\begin{equation} \label{eq:Fermiscale}
v^2 = - \frac{\mu_u^2 - \frac{\Delta M_\phi^2}{2q(1+2M_\phi^2 / M_{Z'}^2)}}{\frac{g^2 + g^{\prime 2}}{4} + \frac{g_x^2}{4(1+M_{Z'}^2/2M_\phi^2)}} \, .
\end{equation}
This means that $\Delta M_\phi^2$ introduces a finetuning in $v^2$ at tree level:
$$
\Delta_v = \left| \frac{\partial \log v^2}{\partial \log \Delta M_\phi^2} \right| = \left| \frac{\Delta M_\phi^2}{v^2} \, \, \frac{1}{\frac{g^2 + g^{\prime 2}}{4} + \frac{g_x^2}{4(1+M_{Z'}^2/2M_\phi^2)}} \, \, \frac{1}{2q(1+2M_\phi^2 / M_{Z'}^2)}\right| \,
$$ 
which however is not a stringent bound. In fact if we tolerate $\Delta_v=10$ then we can have $\Delta M_\phi^2$ up to about $(1 \mbox{ TeV})^2$ in the interesting region of the parameter space, that is when $m_h$ is maximized. It is easy to see that, assuming $\Delta M_\phi^2=0$ at the scale $M$, the running typically generates much smaller splittings. More precisely one finds, up to two loops, neglecting the Yukawa couplings and gaugino soft masses:
\begin{eqnarray}
\frac{d \Delta{M}^2_{\phi}}{d \log \mu} = \frac{4 g_x^2 q}{16 \pi^2} \left[q \Delta M_\phi^2 + \sum_{j \in MSSM} q_j m_j^2 \right] - \frac{4 g_x^2 \lambda^2 q^2}{(16 \pi^2)^2} \Delta M_\phi^2 \label{beta:DeltaMphi}\\
+ \frac{16 g_x^4 q}{(16 \pi^2)^2} \left[ q^3 \Delta M_\phi^2 + \sum_{j \in MSSM} q_j^3 m_j^2 \right] . \nonumber
\end{eqnarray}
Here and in the following we make use of the results of \cite{Martin:1993zk}.
We immediatly see that if there is complete degeneracy at the scale $M$, ie $M_{(\phi)} = M_{(\phi^c)}$ and equal soft masses for the $1^{st}$ and $2^{nd}$ generation sfermions (the only ones which can be large enough to be relevant), then the running of $\Delta M_\phi^2$ starts beyond the two loop level. Thus, with this degeneracy assumption, we can safely neglect $\Delta M_\phi^2$ in all our considerations.

Let us consider the implications on the maximum value for $m_h$, equation (\ref{boundHiggsmass}). 
The largest allowed Higgs boson mass is realized when we include (\ref{boundOnMphi}) in (\ref{boundHiggsmass}), so that we obtain:
\begin{equation} \label{BoundMhWithFinetuning}
m_h^2 \leq \left( m_z^2 +  \frac{g_x^2 v^2}{2(1+\frac{2 q^2 g_x^2}{\lambda^2 \Delta_u})} \right) \, .
\end{equation}
This means that, to increase $m_h$ as much as possible, we prefer a large $\lambda$.
The only problem is then the possibility of a Landau pole, however we see that this can be avoided. The running of $\lambda$ is given by:
$$
\beta_\lambda(\mu) = \left\{
\begin{array}{ll}
\frac{1}{16\pi^2} \left[ 3\lambda^2 - 4 g_X^2 q^2 \right]  & \mbox{ if } \mu >10 \mbox{ TeV} \\
0 & \mbox{ if } \mu <10 \mbox{ TeV}
\end{array}
\right.
$$
where $10$ TeV is an estimate of the scale of the soft masses $M_s$ and $M_\phi$. We write $g_X$ instead of $g_x$, with $g_X(200 \mbox{ GeV})=g_x$, anticipating the notation of Section \ref{sec:runningsU1}.
Thus a sufficient condition to avoid the Landau pole is:
\begin{equation} \label{conditiononlambda}
\lambda^2(200 \mbox{ GeV}) = \lambda^2(10 \mbox{ TeV}) \leq \frac{4}{3} q^2 g_X^2(10 \mbox{ TeV}).
\end{equation}
Notice that at this level there is no substantial difference in $m_h^{max}$ for different values of $q$, the only change coming from the difference in the running of $g_X$ from 200 GeV to 10 TeV which is just a small correction.
However in the following we will see that the interplay between naturalness and EWPT constraints prefers $q=\frac{1}{2}$.

\begin{figure}[thb]
\begin{center}
\begin{tabular}{cc}
\includegraphics[width=0.43\textwidth]{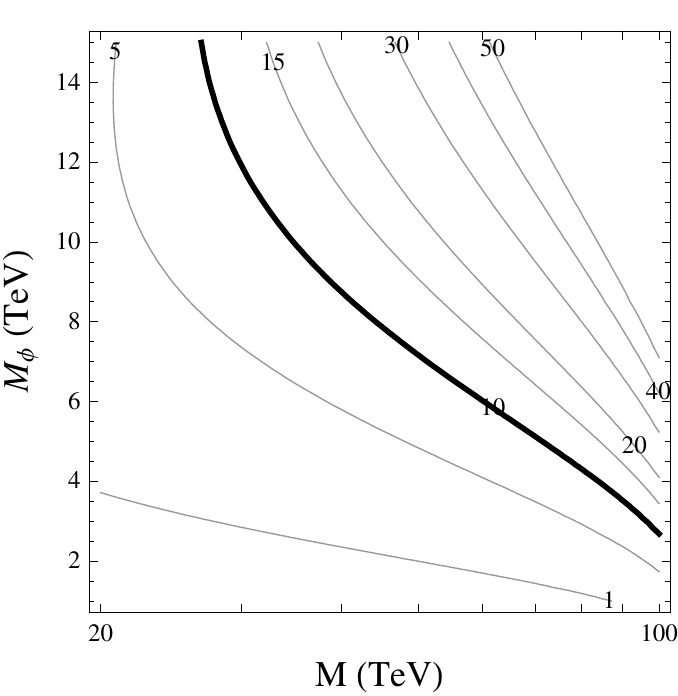} &
\includegraphics[width=0.43\textwidth]{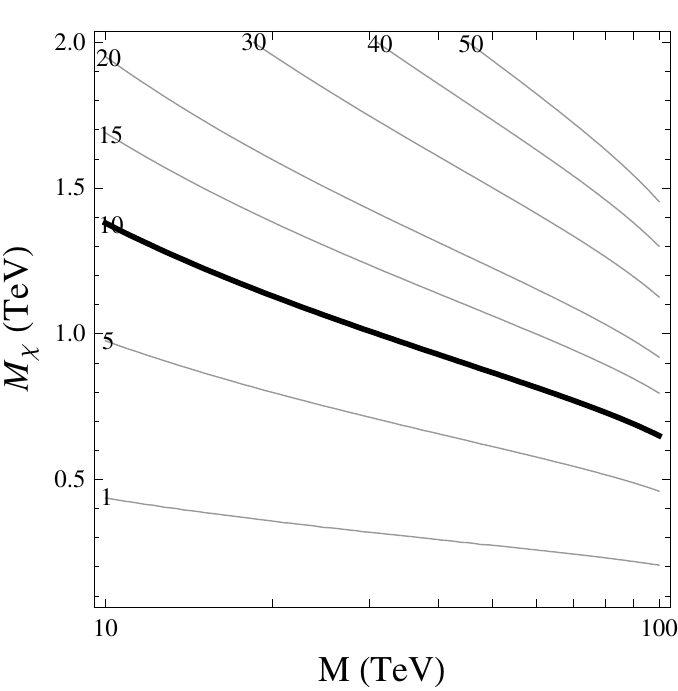}
\end{tabular}
\end{center}
\caption{\small{Finetuning $\Delta$ (\ref{eq:finetuning:Fermiscale}), as a function of the scale $M$ and the soft mass parameter ($M_\phi$ on the left for $q=\frac{1}{2}$, $M_\chi$ on the right), for $g_x(\mbox{200 GeV})=1.3$ ie $m_h^{max}= 2m_Z$, saturating (\ref{boundOnMphi}). The thick line stands for $\Delta =10$.}}
\label{figura:DeltaMphiMchi}
\end{figure}

We now turn to the finetuning at loop level from $M_\phi$, again neglecting the contributions from Yukawa couplings and gaugino soft masses.
From (\ref{eq:Fermiscale}) we see that, if we allow an amount of finetuning $\Delta$, then the radiative corrections to $m_{H_u}^2$ have to satisfy:
\begin{equation} \label{eq:finetuning:Fermiscale}
\delta m^2_{H_u} \leq \, \left( \frac{m_Z^2}{2} + \frac{g_x^2 \, v^2}{4(1+M_{Z'}^2/2M_\phi^2)} \right) \times \Delta
= \frac{(m_h^{max})^2}{2} \times \Delta 
\end{equation}
instead of the usual $\Delta \times \, \frac{m_Z^2}{2}$, as can be seen from (\ref{mzvevsLargeTanBeta}). Neglecting $\Delta M_\phi^2$ one finds:
$$
\frac{d m^2_{H_u}}{d t} =   \frac{4 g_x^4 q^2}{(16 \pi^2)^2}  {M}_{\phi}^2 
\, .
$$
Taking into account the running of $g_x$ only, the result is shown in Figure \ref{figura:DeltaMphiMchi} (left), for $q=\frac{1}{2}$ and $g_x(\mbox{200 GeV})=1.3$, which corresponds to $m_h= 2m_Z$ for large $\tan \beta$ after saturating (\ref{boundOnMphi}) and (\ref{conditiononlambda}).
The lines represent the correction $\delta m_{H_u}^2$ due to $M_\phi^2$ in terms of $\Delta$, as defined in (\ref{eq:finetuning:Fermiscale}). The kinetic mixing effects discussed in Section \ref{sec:runningsU1} are neglected at this stage, since they are just a small correction.

On the other hand from the loop involving the gaugino $\tilde{\chi}$ we have:
\begin{equation} \label{1loopcorrMchi}
\frac{d m_{H_u}^2}{d \log \mu} = - \frac{ 2 g_x^2}{16\pi^2}  \, M_{\chi}^2 \, .
\end{equation}
where $M_\chi$ is the soft mass term in (\ref{softterms}).
In Figure \ref{figura:DeltaMphiMchi} (right) we report the analogous bound on $M_\chi$. 
As already said, however, a small $M_\chi$ does not mean that there is a light particle.

\subsection{Running of gauge couplings and kinetic mixing}
\label{sec:runningsU1}

In general in the presence of two $U(1)$ gauge groups, the Lagrangian contains a mixing:
$$
\mathcal{L}_{gauge} = -\frac{1}{4}{F^{(b)}}_{\mu\nu} {F^{(b)}}^{\mu\nu} -\frac{1}{4}{F^{(a)}}_{\mu\nu} {F^{(a)}}^{\mu\nu} + \frac{\alpha}{2} {F^{(a)}}_{\mu\nu}{F^{(b)}}^{\mu\nu}
$$
whose Feynman rule is (with momenta $k^{\mu}$ and $k^\nu$ in the external legs):
$$
-i\mathcal{M} = i \alpha (k^\mu k^\nu - k^2 g^{\mu\nu}) \, .
$$
On the other hand, if not already present, this term will be generated by radiative corrections. In fact the one loop polarization amplitude connecting the two gauge bosons involving a chiral superfield with charges $q_a$ and $q_b$ is given by:
$$
-i\Pi^{\mu\nu}(k) =  i (k^\mu k^\nu - k^2 g^{\mu\nu}) \frac{g_a g_b}{16 \pi^2} q_a q_b \log \frac{\mu}{\mbox{mass}}
$$
which means (for small $\alpha$):
\begin{equation} \label{RGEalpha}
\frac{d\alpha}{d t} =  \frac{2 g_a g_b}{16 \pi^2} \mbox{Tr}[Q_a Q_b] + o(\alpha) \, .
\end{equation}
Let us see which are the phenomenological consequences.
The kinetic term can be diagonalized with the redefinition:
$$
\left\{
\begin{array}{l}
V^{(a)}_\mu \rightarrow {V^{(a)}}'_\mu + \alpha {V^{(b)}}'_\mu \\
V^{(b)}_\mu \rightarrow {V^{(b)}}'_\mu 
\end{array}
\right.
$$
so that the new charges are, respectively:
$$
\begin{array}{lcl}
g_a Q_a  & \rightarrow & g_a Q_a \\
g_b Q_b & \rightarrow & g_b Q_b + \alpha g_a Q_a \, .
\end{array}
$$
This means that, instead of speaking about kinetic mixing, we can just use three gauge couplings and automatically diagonal kinetic terms. In our case we have $Q_a = Y$ and $Q_b = Y + X$ (see Table \ref{cariche}), so that diagonalizing away the kinetic mixing amounts to transform:
$$
\begin{array}{lcl}
g' Y  & \rightarrow & g' Y \\
g_x (Y+X) & \rightarrow & (g_x + \alpha g' )Y + g_x X \, .
\end{array}
$$
Thus in general we can redefine the model by saying that the coupling of the new vector, in the basis with diagonal kinetic terms, is $g_Y Y + g_X X$. Then we can impose $g_X = g_Y = g_x$ at low energies, and everything is fixed. This is enough for our purposes. The RGE can be taken from \cite{Salvioni:2009jp}, and are:
\begin{eqnarray*}
\frac{d g'}{dt} &=& \frac{1}{16\pi^2} b_{YY} g^{\prime 3}   \\
\frac{d g_X}{dt} &=& \frac{1}{16\pi^2}\left( b_{XX}g_X^3 + 2b_{YX} g_X^2 g_Y + b_{YY} g_X g_Y^2    \right)    \\
\frac{d g_Y}{dt} &=& \frac{1}{16\pi^2}\left( b_{YY} g_Y (g_Y^2 + 2 g^{\prime 2}) + 2 b_{YX} g_X (g_Y^2 + g^{\prime 2}) + b_{XX} g_X^2 g_Y    \right)
\end{eqnarray*}
where $b_{Q_a Q_b} = \mbox{Tr}[Q_a Q_b]$.
Notice that these equations are consistent with (\ref{RGEalpha}), since for small $\alpha$:
$$
\left. \frac{d (g_Y- g_X)}{dt}\right|_{g_X = g_Y} = {g'} \, \left. \frac{d \alpha}{dt} \right|_{\alpha=0} \,
$$
as it should be (since $b_{Y \, X+Y} = b_{YY} + b_{YX}$).

In general, for our purposes, the charge of the new vector can be:
$$
Q=Y + \gamma \frac{L-B}{2} + X_\phi
$$
with arbitrary $\gamma$. Since we want the new gauge coupling to grow with energy as less as possible, we should choose the value of $\gamma$ which minimizes:
$$
b_{QQ} = 2 q^2 + 7 + 4 (\gamma-1)^2
$$
and we see that our choice $\gamma = 1$ (or equivalently $Q = T_3^R + X_\phi$) was the optimal one.

\begin{figure}[thb]
\begin{center}
\begin{tabular}{cc}
\includegraphics[width=0.43\textwidth]{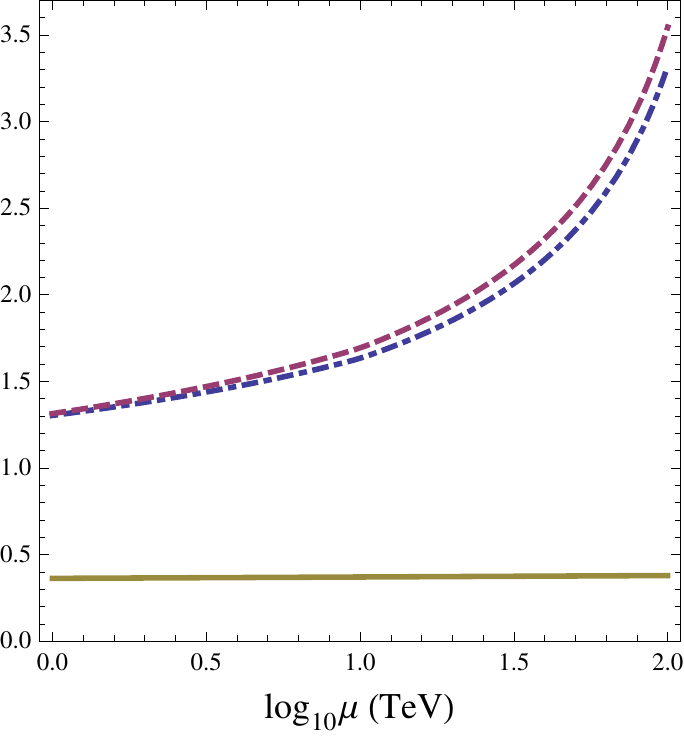} &
\includegraphics[width=0.43\textwidth]{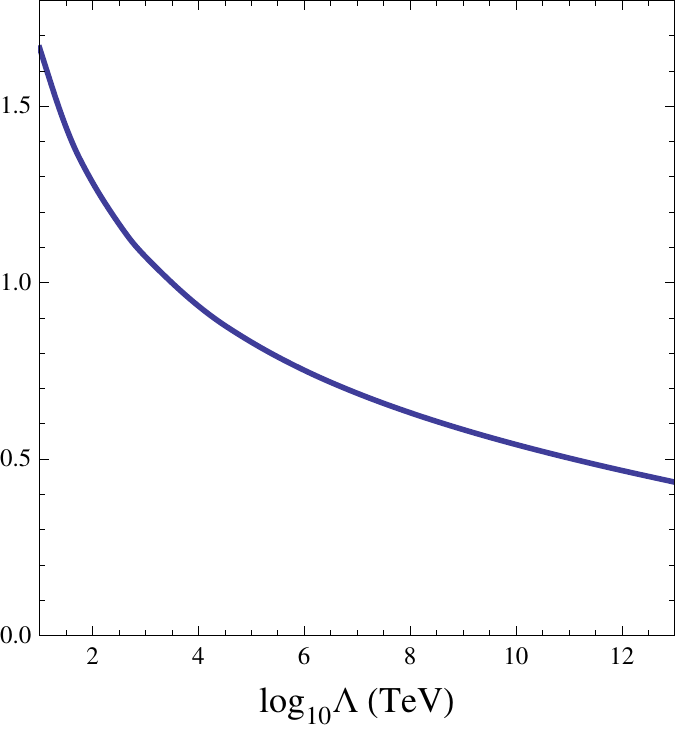}
\end{tabular}
\end{center}
\caption{\small{Left: Running of $g'$ (solid), $g_X$ (dotdashed) and $g_Y$ (dashed) for $q=\frac{1}{2}$ with $g_Y(\mbox{100 TeV})=\sqrt{4\pi}$ and $g_Y(\mbox{200 GeV})=g_X(\mbox{200 GeV})$. Right: value of $g_X=g_Y$ at $200$ GeV versus the scale $\Lambda$ of semiperturbativity ($\alpha_Y(\Lambda)=1$), for $q=\frac{1}{2}$.}}
\label{figura:gxgyRun}
\end{figure}

The values of the coefficients for our model are:
$$
b_{XX} = \left\{\begin{array}{ll} 2q^2 + 4 & \mbox{ if $\mu >$ 10 TeV} \\ 4 & \mbox{ $ \mu <$ 10 TeV} \end{array}  \right. \, , \, b_{YY}=11 \, , \, b_{YX} = -4 \, ,
$$
while the MSSM particles can be effectively decoupled below 200 GeV.
An example of the running is shown in Figure \ref{figura:gxgyRun} (left) if the scale $\Lambda$ at which the model becomes semiperturbative ($g_Y(\Lambda)=\sqrt{4\pi}$) is taken to be $100$ TeV. Notice that $g_Y$ increases faster than $g_X$ because of the kinetic mixing. 
In Figure \ref{figura:gxgyRun} (right) the value of $g_Y(\mbox{200 GeV})=g_X(\mbox{200 GeV})=g_x$ is reported versus the scale $\Lambda$ of semiperturbativity.

Notice finally that the kinetic mixing is not a source of any particular problem or complication, as feared in previous analyses.

\subsection{Experimental bounds} \label{sect:expbounds:U1}

Let us see which are the experimental constraints on this model. The main signature would be that of a $Z'$ boson, which can be extracted from \cite{Carena:2004xs}\cite{Kumar:2006gm}\cite{Contino:2008xg}\cite{Erler:2009jh}\cite{Salvioni:2009mt}.
In the more recent \cite{Salvioni:2009mt} an updated analysis is performed of the present indirect bounds coming from EWPT including LEP2, Tevatron direct searches, and other experiments. Our case corresponds to their Figure 2 with:
$$
\tilde{g}_Y = \frac{g_x}{\sqrt{g^2 + g^{\prime 2}}} \quad , \quad \tilde{g}_X = -\frac{g_x}{2\sqrt{g^2 + g^{\prime 2}}} \, .
$$
For example to get $m_h^{max} \approx 2 m_Z$ we need $g_x = 1.27$, which implies $\Lambda \leq$ 100 TeV as can be seen from Figure \ref{figura:gxgyRun} (right). This means:
$$
\tilde{g}_Y = 1.72 \quad , \quad \tilde{g}_X = -0.86 \, .
$$
which corresponds  to  $M_{Z'}>4-5$ TeV at 95 \% cl. This general analysis is actually performed with $m_h=120$ GeV, however this does not significantly change the final result. In conclusion the model is defendable provided that $M_{Z'}\gtrsim 5$ TeV\footnote{We thank A. Strumia for help on checking this point.}.

Consider now the ratio between $M_\phi$ and $M_{Z'}$. Using (\ref{conditiononlambda}) in (\ref{boundOnMphi}), we find:
\begin{equation*} 
M_{Z'} \geq \sqrt{\frac{3}{10}}  \,  \frac{g_X(\mbox{200 GeV})}{g_X(\mbox{10 TeV})}  \, M_\phi \, .
\end{equation*}
With $m_h = 2m_Z$ we can tolerate $M_{Z'} = 0.40 \, M_\phi$.
Thus we need at least $M_\phi \gtrsim 10-12$ TeV in order to be in agreement with data. This is not in contrast with naturalness only if $M \lesssim 30-35$ TeV, as we see from Figure \ref{figura:DeltaMphiMchi} (left).
An important comment is in order: this is the only point which strongly requires $q=\frac{1}{2}$ instead of $q=1$. The reason is that with $q=1$ the mentioned naturalness bound on $M_\phi$ becomes much more stringent, so that it is difficoult to satisfy it while allowing a sufficiently heavy $Z'$ boson. In fact to allow $M_\phi \gtrsim 10$ TeV we would need $M\lesssim 15$ TeV, which starts being uncomfortably small. For this reason we stick to $q=\frac{1}{2}$.

\subsection{Conclusions - $U(1)$}

With a scale of semiperturbativity $\Lambda \lesssim 100$ TeV and an input scale $M \lesssim 35$ TeV we can have a supersymmetric extension of the Standard Model in which $m_h$ can be as large as $2 m_Z$ at tree level with no more than 10 \% finetuning. This is possible if we allow a large $U(1)_x$ gauge coupling.
The constraints come from: i) naturalness, i.e. Figure \ref{figura:DeltaMphiMchi}; ii) EWPT, which require $M_{Z'}\gtrsim 5$ TeV.
Limitations do not come from $\Delta M_\phi^2 \ll M_\phi^2$ or from the kinetic mixing of $Y$ and $X$, as argued in previous analyses.

\section{Gauge extension $SU(2)$} \label{sect:SU2}

The next-to-minimal version of the model outlined in Section \ref{sect:U1} consists in the addition of a new $SU(2)$ gauge group.
This model is studied in \cite{Batra:2003nj} in a non-universal version in which the gauge interactions of the third generation are different from those of the first and the second ones. The reason is that in that case the new gauge sector is asymptotically free.
However in our bottom-up approach this is just an unnecessary complication, since we are open to changes of regime at intermediate scales. In other words, the point of view we are adopting is to constrain these models through the interplay between naturalness and EWPT, and \textit{not} by the reqirement of unification or perturbativity up to $M_{GUT}$. In any case, we will see that the former constraints are stronger: $M$ is typically required to be much smaller than $\Lambda$  because the running is not so violent, so that our conclusions would basically not change in a non-universal model.

\subsection{Description of the model}

We follow the same line of reasoning of Section \ref{sect:U1}.
To the MSSM we add an extra $SU(2)_{II}$ gauge group in addition to the $SU(2)_{I} \times U(1)_Y$. All the SM fields are charged only under $SU(2)_I$. We also add a $(2,2)$ called $\Sigma$ and a singlet $s$. The transformation law is:
$$
\Sigma \rightarrow U_1 \Sigma U_2^+ \quad , \quad \Sigma =
\left(
\begin{array}{cc}
a & b \\
c & d
\end{array}
\right) \, .
$$
The superpotential is:
$$
W = \lambda \, s \, (a d - bc - w^2) \, .
$$
with soft terms:
\begin{eqnarray} 
\mathcal{L}_{soft} &=& -M_s^2 |s|^2 - M_{\Sigma}^2(|a|^2 + |b|^2 + |c|^2 + |d|^2) \label{softtermsSU2} \\
&& - M_I  \tilde{\chi}_i \tilde{\chi}_i - M_{II}  \tilde{\eta}_j \tilde{\eta}_j + B_s ( ad - bc + h.c.) \, . \nonumber
\end{eqnarray}
It can be seen that all the new fermions take a mass which is controlled by the new breaking scale $u$.
The scalar potential that we have to study is $V =  V_{H\Sigma} + V_\Sigma$ with:
\begin{eqnarray*}
V_{H \Sigma} &=& \mu_u^2 |H_u|^2 + \mu_d^2 |H_d|^2 + \mu_3^2 (H_u H_d + h.c.) \\
&&  + \frac{1}{2} g^{\prime 2}\left(  \frac{1}{2} |H_u|^2 - \frac{1}{2} |H_d|^2 + ..\right)^2 +\frac{1}{2} g_{II}^2 \sum_a \left(\mbox{Tr}\left[ \Sigma T^a \Sigma^+  \right]\right)^2 \, ,\\
&& +\frac{1}{2} g_I^2 \sum_a \left(\mbox{Tr}\left[ \Sigma^+ T^a \Sigma  \right] +  H_u^+ T^a H_u + H_d^+ T^a H_d + ..\right)^2 \\
V_\Sigma &=& \lambda^2 |ad -bc|^2 - B (ad -bc + h.c.) + M_{\Sigma}^2 (|a|^2 + |b|^2 + |c|^2 + |d|^2) 
\end{eqnarray*}
where we wrote only the Higgs and $\Sigma$ fields in the $D$ terms.
Again, the parameters $\mu_3^2$ and $B$ have been made real and positive through field phase redefinition, and $B = \lambda w^2 + B_s \,$ .
The conditions for stability, CP unbreaking and EM unbreaking are the same as before.
We write the configuration of the fields at the minimum as in the $U(1)$ case, with $\left< \Sigma \right> = \mbox{diag}(u_a, u_d)$.
The potential reduces to:
\begin{eqnarray}
V &=& \mu_u^2 v_u^2 + \mu_d^2 v_d^2 - 2 \mu_3^2 v_u v_d  + \frac{1}{8} g^{\prime 2} [v_u^2 - v_d^2]^2 + \frac{1}{8} g_I^2 [ v_u^2 - v_d^2 -  u_a^2 + u_d^2 ]^2 \label{potentialvevSU2}  \\
&&   + \frac{1}{8} g_{II}^2 [  -  u_a^2 + u_d^2 ]^2  + \lambda^2 u_a^2 u_d^2 - 2 B u_a u_d  + M_{\Sigma}^2 (u_a^2 + u_d^2) \, . \nonumber
\end{eqnarray}
We are interested in the case:
$$
u_a^2, u_d^2 \gg v_u^2, v_d^2 \, \, .
$$ 
so we look for an approximate solution of the form:
$$
<a> = u_a = u + \alpha \quad , \quad <d> = u_d = u -\alpha \quad , \quad \alpha \ll u \, .
$$
Solving perturbatively one obtains, at lowest order:
\begin{eqnarray}
u^2 &=& \frac{B - M_\Sigma^2}{\lambda^2} =  \frac{B_s + \lambda w^2 - M_\Sigma^2}{\lambda^2} \qquad (\Rightarrow \mbox{ must be: } B > M_\Sigma^2)  \nonumber \\
\alpha &=& \frac{1}{u} \, \, \frac{\frac{1}{2} g_I^2(v_u^2 - v_d^2)}{2(g_I^2 + g_{II}^2) + \frac{4 M_\Sigma^2 }{B-M_\Sigma^2}\lambda^2}  \, . \label{eq:alphaSU2}
\end{eqnarray}
Substituting in (\ref{potentialvevSU2}) we see that the change with respect to the MSSM equations is:
\begin{equation} \label{sostituzioniSU2}
g^{\prime 2} + g^2  \longrightarrow  g^{\prime 2} +  g_I^2 \frac{g_{II}^2 + \frac{2M_\Sigma^2 }{u^2} }{g_I^2 + g_{II}^2 + \frac{2 M_\Sigma^2 }{u^2}} \, .
\end{equation}
Putting (\ref{sostituzioniSU2}) in the usual results we find that $\tan \beta$ remains the same, ie (\ref{tanbeta}) holds, while the equation relating $m_z$ to the vevs gets modified, in full analogy with (\ref{mzvevs}):
\begin{eqnarray}
&& \frac{v^2}{2} \left( g^{\prime 2} +  g_I^2 \frac{g_{II}^2 + \frac{2M_\Sigma^2 }{u^2} }{g_I^2 + g_{II}^2 + \frac{2 M_\Sigma^2 }{u^2}}  \right) = \frac{\left| \mu_d^2 - \mu_u^2  \right|}{\sqrt{1 - \sin^2 2\beta}} - \mu_u^2 - \mu_d^2 \, . \label{mzvevsSU2}
\end{eqnarray}

In order to use this result we have to connect the gauge couplings with the $Z$ mass, ie to see what corresponds to $g$.
For the moment we work at zeroth order in $\alpha / u$, ie at zeroth order in $v^2 / u^2$. The covariant derivative of $\Sigma$ is:
$$
D_\mu \Sigma = \partial_\mu \Sigma - i g_I {V_1}_\mu^a  T_1^a \Sigma + i g_{II} {V_2}_\mu^b \Sigma T_2^b \, .
$$
At lowest order $\left< \Sigma \right>$ = diag$(u,u)$, so that we have a mass term:
$$
\mathcal{L}_{mass}^{(0)} =\frac{1}{2} \mbox{Tr}\left[\left( - i \frac{u}{2} \sigma^a (g_I V_1^a - g_{II} V_2^a)_\mu  \right) \times h.c.  \right] \, .
$$
This means that the heavy vectors $X_\mu^a$ and light vectors $W_\mu^a$ are, at lowest order in $v^2 / u^2$:
$$
\left\{
\begin{array}{l}
X^\mu = \frac{g_I V_1^\mu - g_{II} V_2^\mu}{\sqrt{g_I^2 + g_{II}^2}} \\
W^\mu = \frac{g_{II} V_1^\mu + g_I V_2^\mu}{\sqrt{g_I^2 + g_{II}^2}}
\end{array}
\right.
$$
and the mass of the heavy vectors is:
\begin{equation} \label{eq:Xmass}
m_{X}^2 = \frac{g_I^2 + g_{II}^2}{2}\, u^2 + o(\frac{v^2}{u^2})\, .
\end{equation}
On the other hand from the Higgs vevs we have, without the hypercharge:
\begin{eqnarray*}
\mathcal{L}_{mass}^{(1)} &=& \frac{1}{2} \mbox{Tr}\left[\left( - i \frac{v}{2} g_I \sigma^a {V_1^a}_\mu  \right) \times h.c.  \right] \\
&=& \frac{1}{2} \mbox{Tr}\left[\left( - i \frac{v}{2} \frac{g_I g_{II}}{\sqrt{g_I^2 + g_{II}^2}} \, \sigma^a \left( W^a + \frac{g_I}{g_{II}} X^a \right)_\mu  \right) \times h.c.  \right] \, .
\end{eqnarray*}
This is equivalent to saying that the $g$ gauge coupling of the MSSM is:
\begin{equation} \label{gcoupling}
g = \frac{g_I g_{II}}{\sqrt{g_I^2 + g_{II}^2}} \, .
\end{equation}
An important point is that all the MSSM fields have an additional coupling to three nearly degenerate heavy vectors $X_\mu^a$, with $SU(2)_L$-like coupling with strength:
\begin{equation} \label{gXcoupling}
g_X = g \frac{g_I}{g_{II}} \, .
\end{equation}

The usual bound on the Higgs boson mass at tree level can be read from (\ref{mzvevsSU2}) and (\ref{gcoupling}):
\begin{equation} \label{boundHiggsmassSU2}
m_h^2 \leq \frac{v^2}{2}\left( g^{\prime 2} + \eta g^2 \right) \cos^2 2\beta \quad , \quad \eta = \frac{1+ \frac{2 M_\Sigma^2}{u^2} \frac{1}{g_{II}^2}}{1+ \frac{2 M_\Sigma^2}{u^2} \frac{1}{g_I^2 + g_{II}^2}} .
\end{equation}
which coincides with equation (3.3) of \cite{Batra:2003nj}.
Notice that this contribution increases with large $g_I$. However $g_I \gg g_{II}$ also implies that the all the $SU(2)_L$ doublets of the MSSM have a large coupling $g_X$ with the heavy vectors (\ref{gXcoupling}), in potential conflict with the EWPT as we discuss below.

The fact that $\alpha$ in (\ref{eq:alphaSU2}) is nonzero means that the complex $SU(2)_{diag}$ triplet, which is contained in $\Sigma$, takes a small nonzero vev. Thus we expect at tree level a correction to the $\rho$ parameter proportional to $\alpha^2$. The precise computation can be done by keeping $\alpha$ in $|D_\mu \Sigma|^2$ and then diagonalizing the full mass matrix. The result at the lowest relevant order is that $m_Z$ is unchanged while:
$$
m_W^2 = \frac{g^2 \, v^2}{2} + 2 g^2 \alpha^2
$$
which means:
$$
\rho = 1 + 4 \frac{\alpha^2}{v^2} \, .
$$
One can define the triplet mass $M_T = M_{\Sigma}^2 + \frac{1}{2}(g_I^2 + g_{II}^2)u^2$,
in terms of which:
\begin{equation} \label{eq:deltarho}
\Delta \rho = \frac{1}{16} \frac{g_I^4}{g^2} \, \frac{u^2 m_W^2}{M_T^4} \, \cos^2 (2 \beta)
\end{equation}
This correction however is very small, as discussed in Section \ref{sect:expbounds}.

\subsection{Naturalness bounds}

In analogy with (\ref{boundOnMphi}) we now impose:
$$
\Delta_u = \left| \frac{\partial \log u^2}{\partial \log M_\Sigma^2} \right| = \frac{M_\Sigma^2}{B - M_\Sigma^2} \leq 10
$$
so that:
\begin{equation} \label{boundOnMSigma}
\frac{M_\Sigma^2}{u^2}  \leq 10 \lambda^2 \, .
\end{equation}
On the other hand from (\ref{mzvevsSU2}) we obtain, in the limit of large $\tan \beta$:
\begin{equation} \label{eq:FermiscaleSU2}
v^2 = - \frac{4 \mu_u^2 }{ g^{\prime 2} + \eta g^2 } \, .
\end{equation}
with $\eta$ from (\ref{boundHiggsmassSU2}).
This means that, if we allow a finetuning $\Delta$, then the radiative corrections to the soft term $m_{H_u}^2$ have to satisfy:
\begin{equation} \label{eq:finetuning:FermiscaleSU2}
\delta m_{H_u}^2 \leq  \frac{g^{\prime 2} + \eta g^2}{4} \, v^2 \, \times \Delta= \frac{(m_h^{max})^2}{2} \times \Delta
\end{equation}
in full analogy with (\ref{eq:finetuning:Fermiscale}).

As in the $U(1)$ case, we want to avoid a Landau pole for the Yukawa coupling $\lambda$, whose evolution is:
$$
\frac{d \lambda}{dt} = 
\left\{
\begin{array}{ll}
0 & \mbox{ if } \mu < 10 \mbox{ TeV}  \\
\frac{3 \lambda}{16\pi^2}\, [\lambda^2 - g_I^2 - g_{II}^2]  & \mbox{ if }  \mu > 10 \mbox{ TeV} \\
\end{array}
\right.
$$
where 10 TeV is an estimate of $M_\Sigma$ and $M_s$. We will thus impose:
$$
\lambda(200 \mbox{ GeV}) = \lambda(10 \mbox{ TeV}) \leq g_I^2 + g_{II}^2 |_{10 \, TeV} \approx g_I^2 + g_{II}^2 |_{200 \, GeV}.
$$

\begin{figure}[thb]
\begin{center}
\begin{tabular}{cc}
\includegraphics[width=0.43\textwidth]{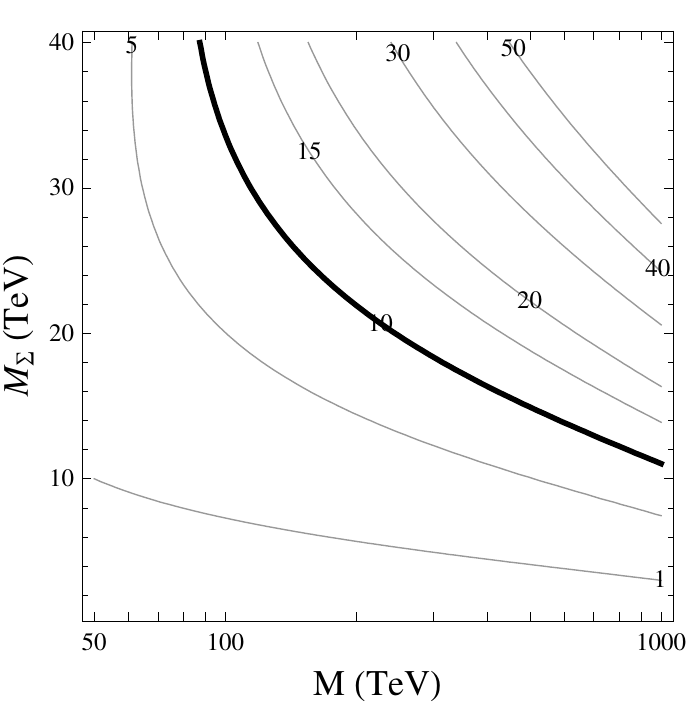} &
\includegraphics[width=0.44\textwidth]{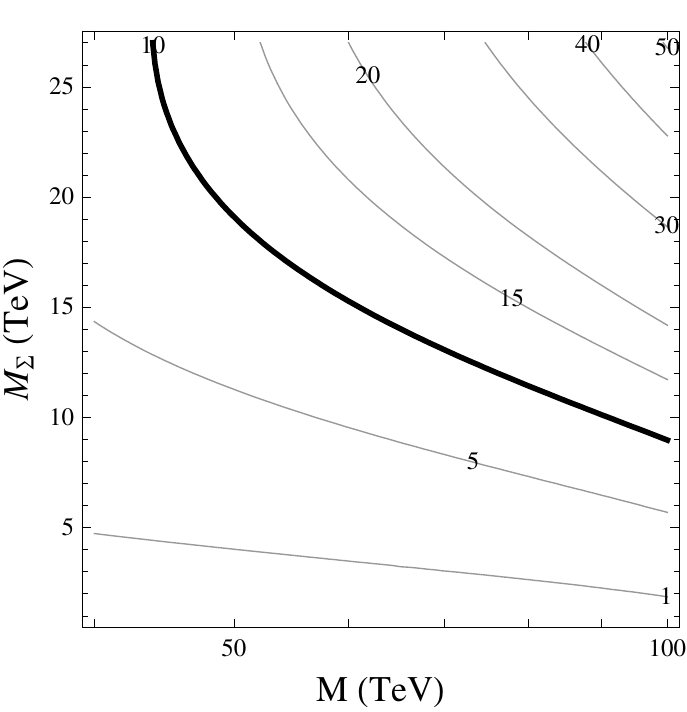}
\end{tabular}
\end{center}
\caption{\small{Finetuning $\Delta$ (\ref{eq:finetuning:FermiscaleSU2}), as a function of the scale $M$ and the soft mass parameter $M_\Sigma$. Left for $m_h^{max}=2m_Z$, right for $m_h^{max}=2.5m_Z$. The thick line stands for $\Delta =10$.}}
\label{figura:DeltaMSigmaSU2}
\end{figure}

We now turn to the radiative corrections to the soft parameter $m_{H_u}$ due to the other soft terms.
At one loop level the only relevant contribution comes from the $SU(2)_I$ gauginos:
\begin{equation} \label{1loopcorrMchiSU2}
\frac{d m_{H_u}^2}{d \log \mu} = - \frac{ 6 g_I^2}{16\pi^2}  \, M_{I}^2 \, .
\end{equation}
where $M_I$ is the soft mass term in (\ref{softtermsSU2}).
Notice again that a low $M_I$ does not imply a low physical gaugino mass, so that this bound is totally irrelevant for our purposes.
The leading contributions coming from $M_\Sigma$ start at two loop order.
To compute it, since $\Sigma$ has no hypercharge, it is sufficient to make the following substitutions in the MSSM formulas:
$$
g_2 \rightarrow g_I \quad , \quad \mbox{Tr}[3 m_Q^2 + m_L^2] \rightarrow \mbox{Tr}[3 m_Q^2 + m_L^2 + 2 M_\Sigma^2] \, .
$$
The result is shown in Figure \ref{figura:DeltaMSigmaSU2}, with the same convention as Figure \ref{figura:DeltaMphiMchi}.
The running of the gauge couplings has been taken into account at one loop level:
\begin{eqnarray} 
\frac{d g_I}{d\log \mu} &=& 
\left\{
\begin{array}{ll}
\frac{1}{16\pi^2} \, \, g_I^3 & \mbox{ if 200 GeV} < \mu < 10 \mbox{ TeV}  \\
\frac{1}{16\pi^2}\, \,  2 \, \,  g_I^3  & \mbox{ if }  \mu > 10 \mbox{ TeV} \\
\end{array}
\right. \label{eq:Rung1g2} \\
\frac{d g_{II}}{d\log \mu} &=& 
\left\{
\begin{array}{ll}
-\frac{1}{16\pi^2}  \, \,  6 \, \,  g_{II}^3 & \mbox{ if 200 GeV} < \mu < 10 \mbox{ TeV}  \\
-\frac{1}{16\pi^2}  \, \,  5 \, \,   g_{II}^3  & \mbox{ if }  \mu > 10 \mbox{ TeV} \\
\end{array}
\right. \nonumber
\end{eqnarray}
where 10 TeV is an estimate of the soft mass of the bidoublet $M_\Sigma$.
Thus perturbativity is not a stringent problem in this model, since the running is much less violent than in the $U(1)$ case.

\subsection{Experimental bounds} \label{sect:expbounds}

The main new feature is now that all the MSSM particles charged under $SU(2)_L$ have also a coupling $g_X$ (\ref{gXcoupling}) to three additional heavy vectors $X^a_\mu$ with mass $m_X$ (\ref{eq:Xmass}). 
With respect to the case of a single $Z'$, this $W'$ case involves in general much more parameters \cite{Rizzo:1996ce}\cite{Rizzo:1996pc}\cite{Weiglein:2004hn}.
Assuming SM-like couplings, Tevatron direct searches exclude a mass below 720-780 GeV \cite{Abachi:1995yi}\cite{Affolder:2001gr}, see also \cite{Abazov:2006aj}.
The LHC is expected to be able to discover heavy charged bosons up to mass of $5.9$ TeV \cite{Rizzo:1996ce}, see also \cite{Cvetic:1993ska}.
The complementary search for $W'$ at $e^+ e^-$ colliders is studied in \cite{Godfrey:2000hc}\cite{Godfrey:2000pw}\cite{Yue:2008jt}.
On the other hand, indirect searches extracted from leptonic and semileptonic decays and from cosmological and astrophysical data give a very wide range of upper limits on $m_{W'}$, depending on the various assumptions and varying from 500 GeV to 20 TeV \cite{Amsler:2008zzb}.

We will impose the relatively safe bound:
\begin{equation} \label{SU2BoundEstimate}
\frac{m_X}{5 \mbox{ TeV}} \gtrsim \frac{g_X}{g_Z} \, .
\end{equation}
Since we stick to $\lambda^2 = g_I^2 + g_{II}^2$ at low energy, for finetunig considerations from (\ref{boundOnMSigma}) we deduce:
\begin{equation} \label{eq:MSigmaMX}
m_X \geq \frac{1}{\sqrt{20}} M_\Sigma \approx 0.22 M_\Sigma \, .
\end{equation}
We are actually interested in the case in which the equality holds, in order to maximize $m_h^{max}$.
For example, if $m_h^{max}=2 m_Z$ ($2.5m_Z$) [$3 m_Z$] then (\ref{SU2BoundEstimate}) gives $m_X \gtrsim 8.5$ (11) [14] TeV, so that (\ref{eq:MSigmaMX}) implies $M_\Sigma \gtrsim$ 40 (50) [60] TeV; from bounds like those in Figure \ref{figura:DeltaMSigmaSU2} we see that naturalness then implies $M<100$ $(\approx50)$ $[\approx60]$ TeV.
Notice that the coupling $g_X$ remains below $\sqrt{4\pi}$ also for very large $m_h$.
However beyond $m_h = 200$ GeV the naturalness bound actually implies $M_\Sigma \approx M$, which means no running at all. In other words, the only possibility in order to be compatible with naturalness becomes to have the new soft scale $M_\Sigma$ very close to the scale $M$, so that the logarithms in the radiative corrections due to $M_\Sigma$ are suppressed. This starts being quite odd, and moreover the main contribution would come from threshold effects which are model dependent. Of course, the situation can be better if we accept a bound less stringent than (\ref{SU2BoundEstimate}).

Notice also that, saturating (\ref{boundOnMSigma}) and using $\lambda^2 = g_I^2 + g_{II}^2$ at low energy, we find $M_T^2 = \frac{21}{20} M_\Sigma^2$ so that (\ref{eq:deltarho}) becomes, for large $\tan \beta$:
$$
\Delta \rho \quad = \quad  \frac{1}{16} \, \frac{g_I^4}{g^2 (g_I^2 + g_{II}^2)} \, \frac{m_W^2}{M_\Sigma^2} \, \left( \frac{20}{21} \right)^2 \, \frac{1}{10} \quad \approx \quad \frac{1}{176} \, \frac{g_I^2}{g_{II}^2} \, \frac{m_W^2}{M_\Sigma^2} \, .
$$
A positive contribution to the $\rho$ parameter in principle would be welcome, since for $m_h \gtrsim 200$ GeV we start being outside of the $2\sigma$ line in the $S-T$ plane of the EWPT fit \cite{:2005ema}. It is in fact true in general that a positive extra contribution to $T$ is helpful in case of a large Higgs boson mass \cite{Peskin:2001rw}.
Unfortunately, with $M_\Sigma$ of the order of 40 TeV, we get $\Delta \rho \sim 10^{-7}$ which is totally negligible.
Thus the case $m_h \geq 2.5m_Z$ is outside of the 95 \% c.l. region in the $S-T$ plane, and one should look for some extra contributions to $T$ in order to defend this possibility.

\subsection{Conclusions - $SU(2)$}

With an input scale $M \sim 100$ (50) TeV we can have a supersymmetric extension of the Standard Model in which $m_h$ can be as large as $2 m_Z$ ($2.5 m_Z$) at tree level with 10 \% finetuning at most. The theory is perturbative up to $\Lambda \backsim 10^{8}$ ($10^3$) TeV. Beyond $m_h=200$ GeV the interplay between naturalness and EWPT starts disfavouring the model, requiring $M_\Sigma \approx M$.
The constraints come from: i) naturalness, ie Figure \ref{figura:DeltaMSigmaSU2}; ii) EWPT, ie (\ref{SU2BoundEstimate}).
The contribution of the small triplet vev to the $\rho$ parameter at tree level is totally negligible. The case $m_h>200$ GeV needs a positive contribution to the $T$ parameter.
The non universal model does not significantly change the situation.

\section{$\lambda$SUSY} \label{sect:lambdasusy}

This last model, which is the NMSSM \cite{Fayet:1974pd}\cite{Ellis:1988er}\cite{Drees:1988fc} with large coupling, is extensively studied in \cite{Barbieri:2006bg} and \cite{Cavicchia:2007dp} to which we refer for details.
In brief, to the MSSM one adds a gauge singlet $s$ with superpotential:
$$
W = \lambda s H_u H_d
$$
and a soft mass $m_s$. Minimizing the scalar potential one finds, for the mass of the lightest Higgs boson at tree level:
\begin{equation} \label{eq:maxmhlsusy}
m_h^2 \leq m_Z^2 \cos ^2 2 \beta + \lambda^2 v^2 \sin^2 2 \beta \, .
\end{equation}
It can be seen \cite{Barbieri:2006bg} that the model is compatible with the EWPT for low $\tan \beta$, with a preferred value less then 3.
Thus we basically have $m_h^{max} = \lambda v$.

\begin{figure}[thb]
\begin{center}
\begin{tabular}{cc}
\includegraphics[width=0.43\textwidth]{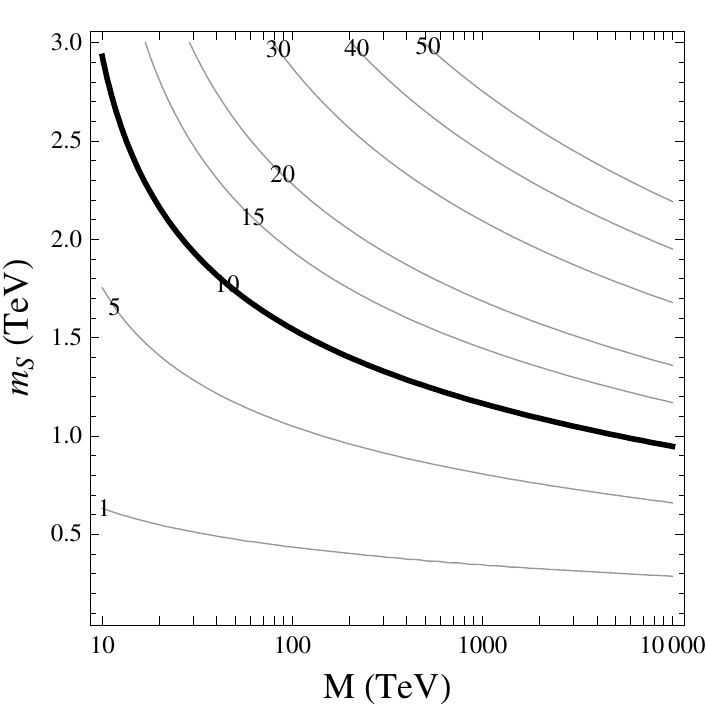} &
\includegraphics[width=0.44\textwidth]{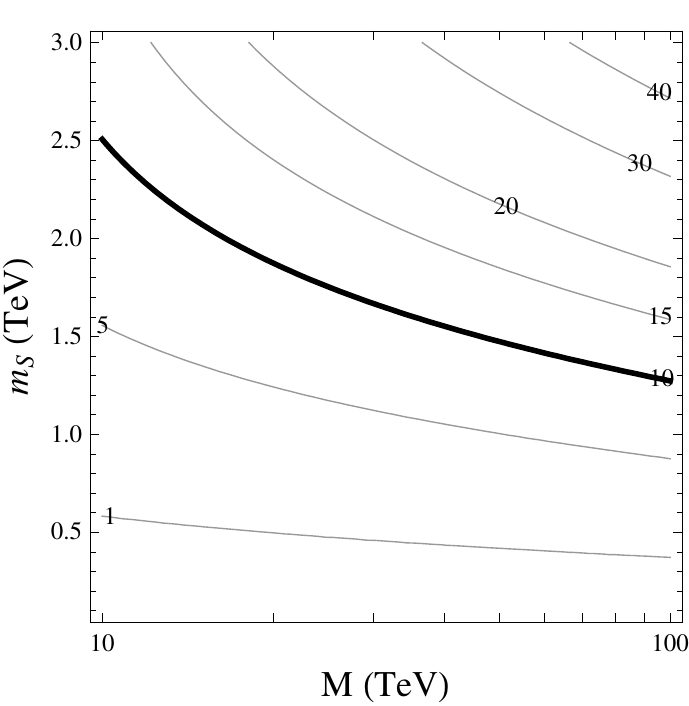}
\end{tabular}
\end{center}
\caption{\small{Finetuning $\Delta$ (\ref{eq:finetuning:FermiscaleLambdaSUSY}), as a function of the scale $M$ and the soft mass parameter $m_s$. Left for $m_h^{max}=2m_Z$, right for $m_h^{max}=3m_Z$. The thick line stands for $\Delta =10$.}}
\label{figura:DeltaMSigmaLambdaSUSY}
\end{figure}

Since we want to increase $m_h$ significantly, the coupling $\lambda$ has to be of order unity at the low scale so that it will typically increase at higher energies.
The relevant RGEs are (see also \cite{Masip:1998jc}\cite{Barbieri:2007tu}\cite{Ellwanger:2009dp}
for more complete analyses):
\begin{eqnarray} \label{exactRGE}
\frac{d \lambda}{d t} &=& \frac{\lambda}{16 \pi^2} \left( 4 \lambda^2 + 3 y_t^2 -3 g_2^2 -g_1^2 \right) \nonumber \\
\frac{d y_t}{d t} &=& \frac{y_t}{16 \pi^2} \left( \lambda^2 + 6 y_t^2 -\frac{16}{3}g_3^2 - 3 g_2^2 - \frac{13}{15}g_1^2 \right). \nonumber 
\end{eqnarray}
For example if $m_h^{max}=2m_Z$ ($m_h^{max}=3m_Z$) then the semiperturbativity scale $\lambda(\Lambda)=\sqrt{4\pi}$ comes out to be $\Lambda \sim 10^4$ TeV ($\Lambda \sim 100$ TeV).
The only extra naturalness constraint is on the soft mass $m_s$. Computing the logarithmic derivative of $v^2$ with respect to $m_s$ one finds (see Section 5.2 of \cite{Barbieri:2006bg}), in the limit in which $\tan \beta =1$, a constraint which is totally analogous to (\ref{eq:finetuning:Fermiscale}) and (\ref{eq:finetuning:FermiscaleSU2}):
\begin{equation} \label{eq:finetuning:FermiscaleLambdaSUSY}
\delta m_{H_u}^2 < \frac{(m_h^{max})^2}{2} \times \Delta \, .
\end{equation}
The bound on $m_s$ which comes from:
$$
\frac{d m_{H_u}^2}{d \log \mu} = \frac{\lambda^2}{8\pi^2} m_s^2
$$
is shown in Figure \ref{figura:DeltaMSigmaLambdaSUSY}, with the same convention as in the other cases. Notice that now we do not have extra experimental constraints which are directly related to the new soft mass, as in the case of the gauge models.

In conclusion, at the price of allowing a large Yukawa coupling $\lambda$ one can significantly increase the masses of the scalar sector of the MSSM consistently with naturalness and EWPT. For example, with semiperturbativity at $10$ TeV the lightest Higgs boson can be as heavy as 350 GeV.
The consequences on the LHC phenomenology are considered in \cite{Cavicchia:2007dp}.

\section{Final remarks} \label{sect:conclusion}

\begin{figure}[thb]
\begin{center}
\includegraphics[width=0.55\textwidth]{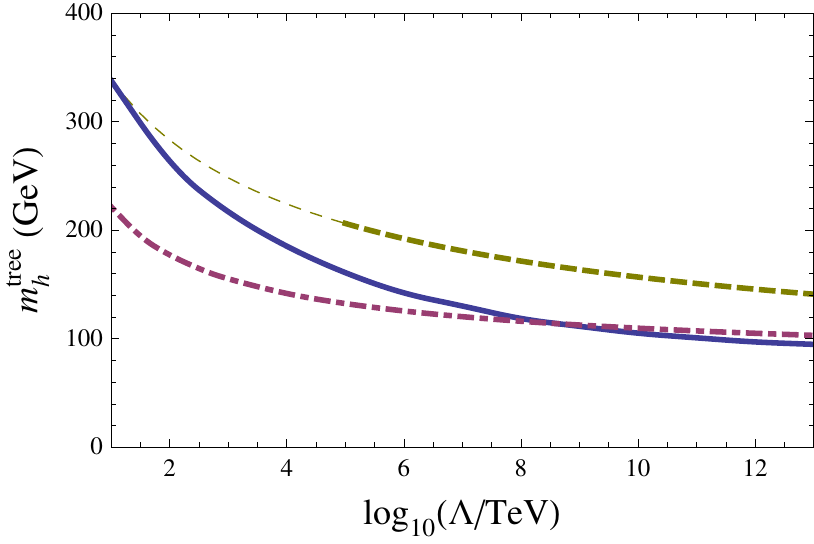}
\end{center}
\caption{\small{Tree level bound on $m_h$ as a function of the scale $\Lambda$ at which $g_I$ or $\lambda$ or $g_X$ equals to $\sqrt{4\pi}$; in the $SU(2)$ model (dashed), in $\lambda$SUSY (solid), and in the $U(1)$ model (dotdashed). For $\lambda$SUSY one needs a low $\tan \beta$, for the gauge extensions one needs a large $\tan \beta$ and 10\% finetuning at tree level in the scalar potential which determines the new breaking scale. In the $SU(2)$ case one should not go much beyond $m_h \sim 200$ GeV, as discussed in Section \ref{sect:expbounds}.}}
\label{FinalFigure}
\end{figure}

We made a comparative study of the three simplest extensions of the MSSM in which the lightest Higgs boson mass can be significantly raised at tree level: a $U(1)$ gauge extension, a $SU(2)$ gauge extension, and $\lambda$SUSY.
From a bottom-up point of view, we discussed the interplay between naturalness and experimental constraints and we showed that the goal can be achieved.
The maximum possible $m_h$ that one can obtain is shown in Figure \ref{FinalFigure} as a function of the scale of semiperturbativity. In the $SU(2)$ case it seems difficoult to be consistent with both the EWPT and naturalness if $m_h$ is beyond 200 GeV.

The prices that one may have to pay are the following: 1) low semiperturbativity scale $\Lambda$; 2) low scale $M$ at which the soft terms are generated; 3) presence of different scales of soft masses; 4) need for extra positive contributions to $T$.
With low scale we mean $\lesssim$ 100 TeV.
With (3) we mean that, besides the usual soft masses of order of hundreds of GeV, one may need some new soft masses of order $10$ TeV.
The ``performance'' of the three models is summarized in Table \ref{table:performance}.
A unified viewpoint on the Higgs mass and the flavor problems for this kind of models will be presented in  \cite{Barbieri:2010pd}.

\begin{table}[thb]
\begin{center}
\begin{tabular}{r|c|c|}
 & $m_h^{max}\, / \,m_Z$ & Price to pay  \\ \hline \hline
$U(1)$ & 2 & (1),(2),(3) \\ \hline \hline
$SU(2)$ & 2 & (2),(3) \\ \hline 
$SU(2)$ & 2.5 & (1),(2),(3),(4) \\ \hline \hline
$\lambda$SUSY & 2 & $-$ \\ \hline
$\lambda$SUSY & 3 & (1) \\ \hline
\end{tabular}
\end{center}
\caption{Summary, see text.}
\label{table:performance}
\end{table}

\section*{Acknowledgements}

I thank Alessandro Strumia for help on Section \ref{sect:expbounds:U1}, Enrico Bertuzzo and Marco Farina for useful comments, and especially Riccardo Barbieri for many important suggestions.

\end{document}